\documentstyle[12pt]{article}

\def\refe{\par\noindent\hangindent=1.5cm}

\begin{document}
\setlength{\tabcolsep}{0.15cm}

\begin{center}
{\large ON THE MAXIMAL QUANTITY OF PROCESSED INFORMATION IN THE
PHYSICAL ESCHATOLOGICAL CONTEXT}

\vspace{1.5cm}

{MILAN M. \'CIRKOVI\'C \\ } {\it Astronomical Observatory,
Volgina 7 \\
11000 Belgrade, Yugoslavia}

\vspace{0.5cm}

{MARINA RADUJKOV \\ } {\it Petnica Science Center, P. O. Box 118
\\ 14000 Valjevo, Yugoslavia}

\end{center}

\vspace{0.5cm}

\begin{abstract}
An estimate of the maximal informational content available to
advanced extraterrestrial or future (post)human civilizations is
presented. It is shown that the fundamental thermodynamical
considerations may lead to a quantitative estimate of the largest
quantity of information to be processed by conceivable computing
devices. This issue is interesting from the point of view of
physical eschatology, as well as general futurological topics,
like the degree of confidence in long-term physical predictions or
viability of the large-scale simulations of complex systems.
\end{abstract}

\section{Introduction}
The problem of intelligent information processing in the
ever-expanding universe has recently been investigated by several
authors (Krauss and Starkman 1999; \'Cirkovi\'c and Bostrom 2000).
These considerations have entered a new phase, building upon
advances in observational cosmology and the strong foundations of
the pioneering studies in cosmological prediction by Rees (1969),
Dyson (1979), Frautschi (1982), Tipler (1986) and Barrow and
Tipler (1986). In the same time, interest in the fundamental
physical limits of {\it computation\/} has grown, largely
motivated by tremendous advances in computer science (e.g.\ Lloyd
2000). Continuation of such trend of technological progress will
lead humanity, sooner or later, into a stage of highly advanced
galactic civilization, in recent years conventionally dubbed the
{\it posthuman\/} era. Apart from the epochal importance of any
such possible development for biological and social sciences, it
has a strong bearing on the nascent astrophysical discipline of
physical eschatology, for the reasons suggested boldly by Dyson
(1979), when he wrote:
\begin{quote}
It is impossible to calculate in detail the long-range future of
the universe without including the effects of life and
intelligence. It is impossible to calculate the capabilities of
life and intelligence without touching, at least peripherally,
philosophical questions.
\end{quote}
Certainly, the most important property of intelligence is its
capacity for information processing. As it has been recognized
for a long time, this form of information processing is still
entirely within limits of physical, specifically thermodynamical,
laws. This conclusion does not necessarily entail reductionism,
since consciousness may still contain uncomputable and therefore
irreducible elements {\it in addition\/} to the conventionally
established ones (e.g.\ Penrose 1989). And thermodynamics in
general is, in the limit of very long timescales, determined by
properties of the universe as a whole, i.e.~by astrophysics and
cosmology. Our goal in this note is to consider very simple cases
of maximal informational resources available to advanced
civilizations, either extraterrestrial or posthuman. We do not
enter into any possible distinction between the two, since the
ages of humanity, Earth and our Galaxy, as well as the pace of
chemical evolution, suggest that there may be intelligent
communities much older than ours even in our cosmological
neighbourhood (Livio 1999).

\section{An estimate of the maximal quantity of processed information}
Let us consider various energy fields as potential sources of
energy for information processing, using the assumption ("Cosmic
Sum rule") that $\Omega = \Omega_\Lambda +\Omega_b + \Omega_{\rm
CDM} = 1$, where the first term corresponds to the vacuum energy,
second to the baryonic matter, and the third to the
(non-baryonic) cold dark matter (for the general cosmological
considerations, see Peebles 1993). The best present estimates
approach values $\Omega_\Lambda \approx 0.7$, $\Omega_{\rm CDM}
\approx 0.25$ and $\Omega_b \approx 0.05$ (e.g.\ \'Cirkovi\'c and
Bostrom 2000, and references therein). One cannot do anything
useful with the vacuum energy.\footnote{On the contrary, it
plausibly harms the energy budget by preventing, through the
so-called cosmological "no-hair" theorem, the development and
subsequent exploitation of the cosmological shear during the
Hubble expansion (Gibbons and Hawking 1977).} As far as CDM is
concerned, it could conceivably be used as an energy source,
since the annihilation of these cosmions and anticosmions
(present in approximately equal numbers according to the standard
theory) would produce potentially usable energy (for some
consequences of annihilation, see Kaplinghat, Knox and Turner
2000). However, depending on the mass spectrum of cosmions, their
galactic density is rather small, and since their interactions
are by definition very weak, their gathering and separation for
annihilation will pose huge engineering problems. If we consider
the model of "posthumanity soon", than only usable matter field
is the baryonic matter, in the local environment concentrated in
the form of planetary systems.

We may use the Brillouen's (1962) equality for the upper limit of
processable quantity of information (in bits):
\begin{equation}
\label{jedan} I_{\rm max} = \frac{\Delta E}{k_B T \ln 2}= 1.05
\times 10^{16} \frac{\Delta E}{T}.
\end{equation}
Here, $k_B$ is the Boltzmann constant. Both $\Delta E$ and $T$
are functions of time, as well as of cosmological parameters
$\Omega$, $\Lambda$ and $H_0$. There are several important
consequences of the Brillouen's inequality for cosmology. The
amount of information available for processing in any causally
connected region of finite proper size with trivial topology is
necessarily finite.

As for the working temperature of posthuman computers, we may
wish to consider two somewhat extreme models: (I) the first with
$T= T_{\rm CMB} = 2.730 \pm 0.014$ K (Staggs et
al.~1996)---indicating quick reaching of posthuman stage;
additional cooling would require diverting the precious energy
resources from computing itself---and (II) the second with
\begin{equation}
\label{spec}
 T = T_{\rm vac} =\frac{\hbar c}{k_B}
\sqrt{\frac{\Lambda}{12 \pi^2}} = 3.2985 \times 10^{-30} h \left(
\frac{\Omega_\Lambda}{0.7} \right)^\frac{1}{2} \; {\rm K}
\end{equation}
indicating posthumanity in the asymptotic limit of physical
eschatology. In this equation, $\Lambda$ is the vacuum
energy-density corresponding to the dimensionless cosmological
density fraction $\Omega_\Lambda$, and $h$ is the dimensionless
Hubble constant ($H_0 \equiv 100\; h$ km s$^{-1}$ Mpc$^{-1}$).

Now we can write for the expendable energy:
\begin{equation}
\label{dva} \Delta E = \int\limits_V q ( \Omega_b \rho_{\rm crit})
c^2 \; dV,
\end{equation}
where $V$ is the proper volume available to posthuman
civilization and $q$ is the energy extraction efficiency (between
0 and 1, and optimistically over 0.5). The usage of $\Omega_b$ is
particularly useful when we consider possibility of posthuman
civilizations of truly intergalactic size. Currently there is
some confusion about the exact value of $\Omega_b$, since the
primordial nucleosynthesis inferences apparently conflicts with
conclusions drawn from the microwave background anisotropy
observations (e.g.\ Kaplinghat and Turner 2001). It is to be
expected that this problem will be solved soon, but in the
meantime we note that the exact value is inessential for our
purposes. If we restrict ourselves to the model (I), that is,
"posthumanity soon", we may neglect spatial variation of $q$ and
$\Omega_b$ (which would be present in other cases, say $q$ is
certainly different in intergalactic space from the one within
galaxies), and using the definitional relation for $\Omega_b$ we
get
\begin{equation}
\label{tri} \Delta E \approx qc^2 \int\limits_V \rho_b \; dV.
\end{equation}
Now, plugging this into (1), we obtain
\begin{equation}
\label{cetiri} I_{\rm max} = \frac{qc^2}{k_B \ln
2}\frac{\int\limits_V \! \rho_b \; dV}{T} = 3.5 \times 10^{36} q
\int\limits_V \rho_b \; dV,
\end{equation}
in bits. If we wish to consider truly short-term posthuman
civilization, we may state that the value of the integral is
equal to
\begin{equation}
\label{pet} \int\limits_V \! \rho_b \; dV = 2 \times 10^{33}
\frac{\bar{M}}{M_\odot} n \; \; {\rm grams},
\end{equation}
where $n =1, 2 ...$ is the number of planetary systems controlled
by our prototype posthuman civilization, and $\bar{M}$ is the
average mass of a planetary system in the Galaxy. The fraction in
(\ref{pet}) is likely to be less or about 1.5, when mass of the
hidden matter (like comets in the Oort cloud whose total mass is
still uncertain; see Weissman 1983) in our planetary system is
taken into account, as well as the possibility of harvesting some
interstellar matter between systems. Taking all this into
account, we reach the estimate of
\begin{equation}
\label{sest} I_{\rm max} = 7 \times 10^{69}
qn\frac{\bar{M}}{M_\odot} \; \; {\rm bits}.
\end{equation}
An analogous estimate can be obtained for the case (II) of
posthumans in the far future. This case will certainly be prone
to much larger uncertainties of not only quantitative, but also
qualitative nature. Therefore, we shall here give just a sketch
and postpone the detailed analysis to a forthcoming study. To this
end, we may generalize the expression (\ref{cetiri}) taking into
account the changes in both resources and methods available to
advanced civilizations during long cosmological scales
\begin{equation}
\label{sedam} I_{\rm max} (t) = \frac{q(t) \, c^2}{k_B \ln 2}
\int\limits_{t_0}^t  \! \frac{dt}{T(t)} \int\limits_{V(t)} \!
\rho_b \; dV.
\end{equation}
Here, $V(t)$ denotes the physical (proper) volume of space
available to the advanced civilization under consideration, and
$T(t)$ is the evolution of the cosmic temperature ($T(t_0) =
T_{\rm CMB} = 2.730 \pm 0.014$ K). In this case, it is important
to notice that the evolution of the equilibrium temperature of
the universe in the context of physical eschatology is a problem
not exactly solved to this day (approximate results for
Einstein-de Sitter model are presented in Adams and Laughlin
1997), mainly because at very late epochs controversial sources
(like the proton decay and Hawking radiation of an uncertain
number of decaying black holes) come into play. Fortunately, we
may be in position to know the asymptotic limit of the process of
cooling of the universe, since it is determined by the temperature
associated with the cosmological event horizon (Gibbons and
Hawking 1977) in eq.~(\ref{spec}). Since it can be shown
that---if current estimates of the cosmological constant are
correct---we have already entered the exponentially expanding
(quasi)de Sitter phase (\'Cirkovi\'c and Bostrom 2000), we may
use the simplest approximation of temperature decreasing as $T
(t) = T (t_0) \exp [-H (t-t_0)]$, leading to the formal
equalisation with (\ref{spec}) in $t = t_0 + (1/H_0) \ln [T (t_0)
/ T_{\rm vac}] \approx t_0 + 69.58 /H_0$, i.e.~in about 1140 Gyr
(for $h=0.6$). As shown by Adams and Laughlin (1997; although only
for the Einstein-de Sitter case), this is too pessimistic, since
other photon sources will replace CMB as the determinants of the
temperature of the universal heat bath. It will take probably
many orders of magnitude larger time for the universe to reach
the de Sitter temperature, but what is important is that there
may be no further cooling. Thus, Dyson's (1979) idea about using
a special form of hybernation for the expression in
eq.~(\ref{jedan}) to diverge is unfeasible. That said, we leave
considerations of the detailed solutions of eq.~(\ref{sedam}) to a
subsequent work, which will benefit from our increased
understanding of the future thermal history of the universe.

\section{Discussion}
We have estimated (within an order of magnitude) the information
processing power of advanced extraterrestrial or future posthuman
communities. Our estimate is conservative in the sense that we
have ignored the complicated and not yet fully understood issue of
dissipationless computation (see, for instance, Porod et al.\
1984) and assumed classical limit of $kT \log 2$ dissipation per
logical step. In addition, we have neglected the important issues
of information {\it transmission\/}, noise and error correction
(for a preliminary treatment of these, see the pioneering study of
Sandberg 2000). All these issues should be covered in future,
more realistic treatments.

It is worth noticing that numerical estimates reached above are
quite conservative in comparison to the proposed fundamental
information bounds, such as the holographic bound (e.g.\ Sussking
1995)
\begin{equation}
\label{holo} I_{\rm max}^h = \frac{A}{4 L_{Pl}^2 \ln 2},
\end{equation}
(where $L_{Pl} \equiv \sqrt{G \hbar /c^3} \approx 2 \times
10^{-33}$ cm is the Planck length) or the Bekenstein (1973, 1981)
bound
\begin{equation}
\label{beken} I_{\rm max}^B = \frac{2 \pi R E}{c \hbar \ln 2},
\end{equation}
where $R$ is the radius and $E$ the total mass-energy of the
information cache. For instance, the equivalent of a holographic
bound for the case of entire planetary systems from eq.\
(\ref{sest}) gives $I_{\rm max}^h \approx 9.8 \times 10^{76} \, (n
\bar{M} / M_\odot)^2$ bits, and the application of the Bekenstein
bound to the same case yields $I_{\rm max}^B \approx 5.13 \times
10^{71} \, q(n \bar{M} / M_\odot) (R / 1 \; {\rm cm})$ bits
(where one should keep in mind that in order for this bound to be
meaningful, the size of the memory $R$ has to be larger from its
gravitational radius, which is about $3 \times 10^5$ cm per Solar
mass). Parenthetically, we note that most of the treatments of
cosmological limitations on computation to be found in the
literature (e.g.\ Tipler 1986; Sandberg 2000) and on the Internet
use the Bekenstein bound, which is certainly more realistic and
practical, but does not look entirely sound from the conceptual
point of view. The reason for such a disadvantage of the
Bekenstein bound compared to the holographic bound is that the
Newtonian gravitational constant $G$ figures explicitly in the
latter and not in the former. Since we believe that quantum
gravitational degrees of freedom may contain a huge amount of
information (and could be potentially exploited by advanced
communities), a liberal approach would favor the usage of the
holographic bound.

There are several reasons for pursuing this problem further. One
is related to the confidence we may have in predictions of
nonlinear dynamical systems by future computing devices. It is a
well-known fact that (apparent) complexity of nonlinear systems
increase at an exponential rate with time. No matter how advanced
computing devices are employed in order to predict the behaviour
of such systems, the prediction will break down at a particular
timescale. One of the main factors determining this timescale is
certainly the maximal amount of information which can be
processed whatsoever during the computations necessary for
prediction. But probably the most interesting issue to be solved
by calculations such as these is the question of information cost
of running large-scale detailed simulations of human environment.
There are reasons to believe that advanced posthuman
civilizations will run such simulations, which are sometimes
aptly called "ancestor simulations" (Prof.\ Nick Bostrom,
manuscript in preparation). It is clear, however, that the depth
(or complexity) of such a simulation would be very sensitive to
the computing power and information resources of a posthuman
"director". In order to assess viability of this scenario, one
should attempt to reasonably predict those quantities. On the
other hand, it remains for computer and cognitive scientists, as
well as sociologists, to answer the deep question of minimal
informational cost of any realistic simulation of human
consciousness and society.

\vspace{0.5cm}

It is a pleasure to express gratitude to Prof.\ Nick Bostrom,
whose deep insights into all sorts of futurological and
philosophical questions has largely motivated the study of this
topic. Srdjan Samurovi\'c is acknowledged for invaluable
technical help.

\vspace{1cm} {\Large \bf \noindent References} \vspace{0.5cm}

\refe Adams, F. C. and Laughlin, G.: 1997, {\it Rev. Mod. Phys.}
{\bf 69}, 337.

\refe Barrow, J. D. and Tipler, F. J.: 1986, {\it The Anthropic
Cosmological Principle\/} (New York: Oxford University Press).

\refe Bekenstein, J.: 1973, {\it Phys. Rev. D\/} {\bf 7}, 2333.

\refe Bekenstein, J.: 1981, {\it Phys. Rev. Lett.} {\bf 46}, 623.

\refe Brillouin, L.: 1962, {\it Science and Information Theory\/}
(New York: Academic Press).

\refe \'Cirkovi\'c, M. M. and Bostrom, N.: 2000, {\it Astrophys.
Space Sci.} {\bf  274}, 675.

\refe Dyson, F.: 1979, {\it Rev. Mod. Phys.} {\bf 51}, 447.

\refe Frautschi, S.: 1982, {\it Science\/} {\bf 217}, 593.

\refe Gibbons, G. W. and Hawking, S. W.: 1977, {\it Phys. Rev.
D\/} {\bf 15}, 2738.

\refe Kaplinghat, M., Knox, L., and Turner, M. S.: 2000, {\it
Phys. Rev. Lett.} {\bf 85}, 3335.

\refe Kaplinghat, M. and Turner, M. S.: 2001, {\it Phys. Rev.
Lett.} {\bf 86} 385.

\refe Krauss, L. M. and Starkman, G.: 2000, {\it Astrophys. J.}
{\bf 531}, 22.

\refe Livio, M.: 1999, {\it Astrophys. J.} {\bf 511}, 429.

\refe Lloyd, S.: 2000, {\it Nature} {\bf 406}, 1047.

\refe Peebles, P. J. E.: 1993, {\it Principles of Physical
Cosmology\/} (Princeton: Princeton University Press).

\refe Penrose, R.: 1989, {\it The Emperor's New Mind\/} (Oxford:
Oxford University Press).

\refe Porod, W., Grondin, R. O., Ferry, D. K. and Porod, G.:
1984, {\it Phys. Rev. Lett.} {\bf 52}, 232.

\refe Rees, M. J.: 1969, {\it Observatory\/} {\bf 89}, 193.

\refe Sandberg, A.: 2000, {\it Journal of Transhumanism\/} {\bf
5} (available at\\ http://transhumanist.com/volume5/Brains2.pdf).

\refe Staggs, S. T., Jarosik, N. C., Meyer, S. S., and Wilkinson,
D. T.: 1996, {\it Astrophys. J.} {\bf 473}, L1.

\refe Susskind, L.: 1995, {\it J. Math. Phys.} {\bf 36}, 6377.

\refe Tipler, F. J.: 1986, {\it Int. J. Theor. Phys.} {\bf 25},
617.

\refe Weissman, P. R.: 1983, {\it Astron. Astrophys.} {\bf 118},
90.

\end{document}